\documentclass{article} % For LaTeX2e
\usepackage{iclr2020_conference,times}
\usepackage{graphicx}
\usepackage{amsmath,amsfonts,bm}

\def\eqref#1{equation~\ref{#1}}
\def\1{\bm{1}}

\DeclareMathAlphabet{\mathsfit}{\encodingdefault}{\sfdefault}{m}{sl}
\SetMathAlphabet{\mathsfit}{bold}{\encodingdefault}{\sfdefault}{bx}{n}

\usepackage{hyperref}
\usepackage{url}

\title{SMArtCast: Predicting soil moisture interpolations into the future using Earth observation data in a deep learning framework}

\author{Conrad J. Foley,  Sagar Vaze, Mohamed Seddiq, Alexey Unagaev, Natalia Efremova \thanks{www.deepplanet.ai} \\
Deep Planet ltd\\
Oxford, UK\\
\texttt{\{James,Sagar,Mohamed,Aleksei,Natalia\}@deepplanet.ai} \\
}

\iclrfinalcopy % Uncomment for camera-ready version, but NOT for submission.
\begin{document}

\maketitle

\begin{abstract}
Soil moisture is critical component of crop health and monitoring it can enable further actions for increasing yield or preventing catastrophic die off. As climate change increases the likelihood of extreme weather events and reduces the predictability of weather, and non-optimal soil moistures for crops may become more likely. In this work, we use a series of LSTM architectures to analyze measurements of soil moisture and vegetation indices derived from satellite imagery. The system learns to predict the future values of these measurements. These spatially \textit{sparse} values and indices are used as input features to an interpolation method that infer spatially \textit{dense} moisture maps at multiple depths for a future time point. This has the potential to provide advance warning for soil moistures that may be inhospitable to crops across an area with limited monitoring capacity.  
 
\end{abstract}

\section{Introduction}

Climate change is one of the most pressing threats globally with the potential to cause frequent or prolonged droughts in many areas of the world \citep{LeHouerou1996}. Infrequent or unpredictable rainfall and higher evapotranspiration due to increased temperature may lead to reduced soil moisture \citep{Kingston2009}. Soil moisture is critical to the growth of almost all arable crops globally impacting small subsistence farmers up to large industrial agricultural companies. Irrigation is required in many regions to maintain suitable soil moisture for crops and is one of the largest fresh water usages in the world \citep{Frenken2012}. Adequate soil moisture is critical for the optimal growth of crops and maximising yield. Non-optimal soil moistures can lead to crop failure which is a threat to the livelihood of farmers and can seriously endanger the robustness of the food supply chain that is vital in the modern world. 

Soil moisture has traditionally been monitored using ground sensors that are buried underground and can report the soil moisture in that soil column \citep{Pagay2016}. These sensors can be extremely costly both in direct equipment cost but also the labour required to embed them down to generally below one meter. These sensors then still come with the draw back that they can only report the soil moisture in that one specific location so a great number of sensors would be required to provide an accurate measure of soil moisture across a large area bringing with it an associated high cost. To avoid this, methods of interpolation have been developed that attempt to estimate the soil moisture between sensors by using algorithms based on values and distance from nearby sensors and other features at the target location \citep{Wahba1990}. Machine and deep learning have become recognised as a tool that can be used to help fight climate change across numerous domains \citep{Rolnick2019}.

\section{Background}

Soil moisture prediction into the future was previously performed through physical models. By taking an input of precipitation and temperature physical models or simple regression models could be used to predict soil moisture using predictions of weather. These models tended to have low accuracy and predict on a low-resolution providing generalisations of soil moisture on a coarse temporal scale. With the advent of machine learning model performance greatly improved and now fine scale climatic predictions such as nowcasting \citep{Agrawal2019} which provides on the spot predictions into very near timescales are possible. But predicting further into the future in a usable way is still a difficult problem. Recently deep learning has become a tool that excels in the field of time series prediction particularly with the use of recurrent neural networks (RNNs) \citep{Rumelhart1986} and long-short term memory RNNs (LSTMs) \citep{Hochreiter1997}. LSTMs have been shown to perform well in prediction of soil moisture values into the future, with increased predictive ability leading to savings in water usage \citep{Adeyemi2018}. For predicting soil moisture not just at sensor locations machine learning frameworks have been applied on satelite imagery using radar or vegetation information \citep{Efremova2019,Abbes2019} but these predictions are only accurate at surface level while many crop species are reliant on deeper soil moistures.

Vegetation health indices such as the normalised difference vegetation index (NDVI) \citep{Rouse1973} and the normalised difference water index (NDWI) \citep{Gao1996} have been used in agriculture and geospatial fields for years. These indices also provide proxy measure of other information about the available water content in plants and soil. Being able to predict these indices in the future as a time series problem would provide information about crop health and other features in the future. Previously NDVI has been predicted using different machine learning frameworks which are able to perform adequately at the task \citep{Das2016,Nay2018,Stas2016}.

Interpolation methods tend to rely on spatial features alone with methods such as inverse distance weighting, splining and kriging all seeing use \citep{Li2011}. These rely on the distance between measurements and provide some form of weights to determine the impact that distance has on the prediction. Combining interpolation methods with soil moisture predictions from LSTM's it becomes possible to predict soil moisture across entire areas into the future with a small number of sensors in a deep learning framework. 

Our aim was to build a pipeline (SMArtCast: \textbf{S}oil \textbf{M}oisture \textbf{Ar}tificial In\textbf{t}elligence fore\textbf{Cast}) that can take soil moisture data from sparse sensors and provide predictions of soil moisture across an entire area up to two weeks into the future at all sensor depths using satellite imagery to supplement features. This would be capable of providing advanced warning of soil moistures that are deleterious to crops as well as to provide insights into the effectiveness or requirements of irrigation leading to the maximisation of crop yields and to possible reduction in freshwater usage. We present evidence that using deep learning methodologies it is possible to predict multiple features into the future and interpolate soil moisture across an entire satellite image.

\section{Materials and Methods}
\label{gen_inst}
%\subsection{Data}
As input data, we use soil moisture measurements installed on 200 hectares of land.  Input measurements are taken from the embedded soil moisture sensors, each installed at 10 centimeters depth up until 120 cm depth; rainfall and temperature data; and high resolution satellite imagery with 13 spectral bands. 

\subsection{Soil Moisture Prediction}
\begin{figure}[!h]
    \centering
    \includegraphics[width = 8cm, height = 4cm]{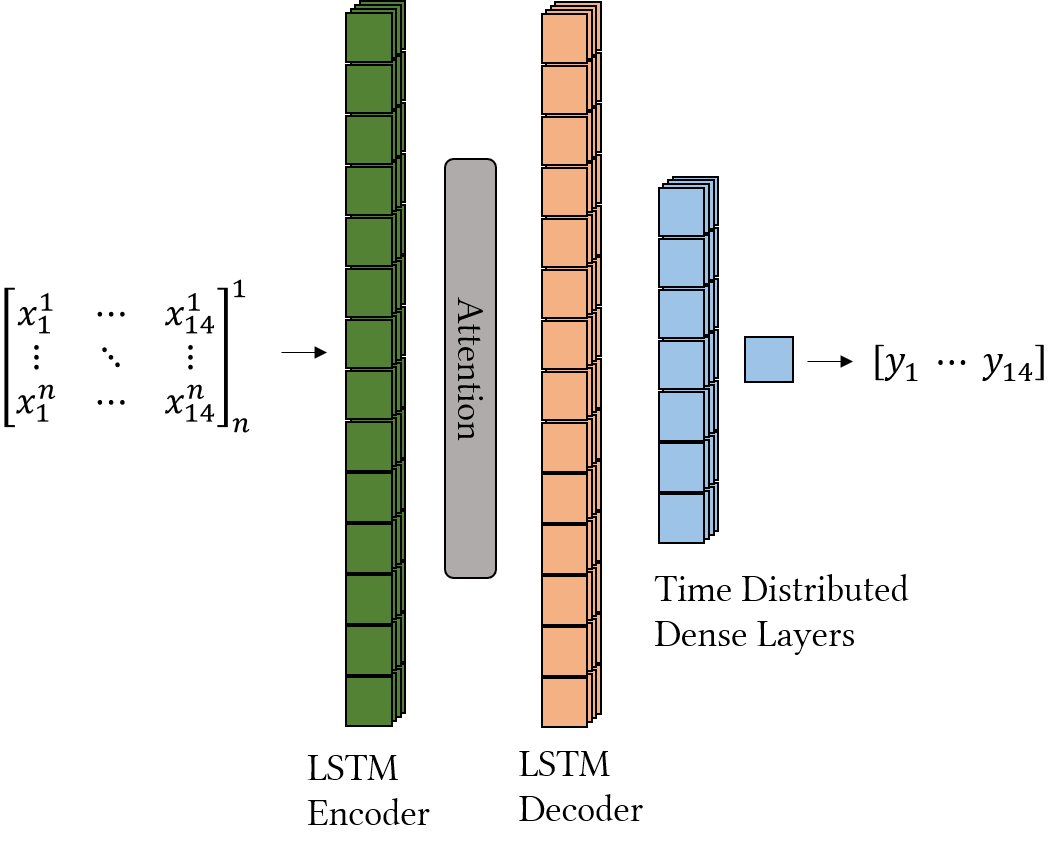}
    \caption{Structure of the LSTM encoder decoder model for soil moisture prediction.}
    \label{fig:SM_lstm}
\end{figure}

The input of the proposed LSTM model is a matrix of features that include soil moisture, soil temperature, soil salinity and rainfall and provides an output of 14 days of soil moisture predictions, 1 per day of predictions. The model architecture is based on sequence to sequence learning using an LSTM encoder - decoder \citep{Sutskever2014}. The encoder layer and decoder layer both had 200 neurons and a \textit{tanh} activation function. The output of the decoder is then passed through a time distributed dense layer with 100 neurons and a final dense layer with a single neuron both with linear activation functions (see Fig. \ref{fig:SM_lstm}).

\subsection{Satellite imagery time series prediction}
Satellite imagery of the study site is acquired and then transformed into the normalised difference vegetation and water indices (NDVI and NDWI). These indices rely on the satellite image bandwidths in the Red, near infrared (NIR) and short-wave infrared (SWIR). 

\begin{equation}
 \begin{aligned}
  NDVI=\frac{NIR-Red}{NIR+Red}, \;\; NDWI= \frac{NIR-SWIR}{NIR+SWIR} 
 \end{aligned}
\end{equation}

\begin{figure}[!h]
    \centering
    \includegraphics[width = 10cm]{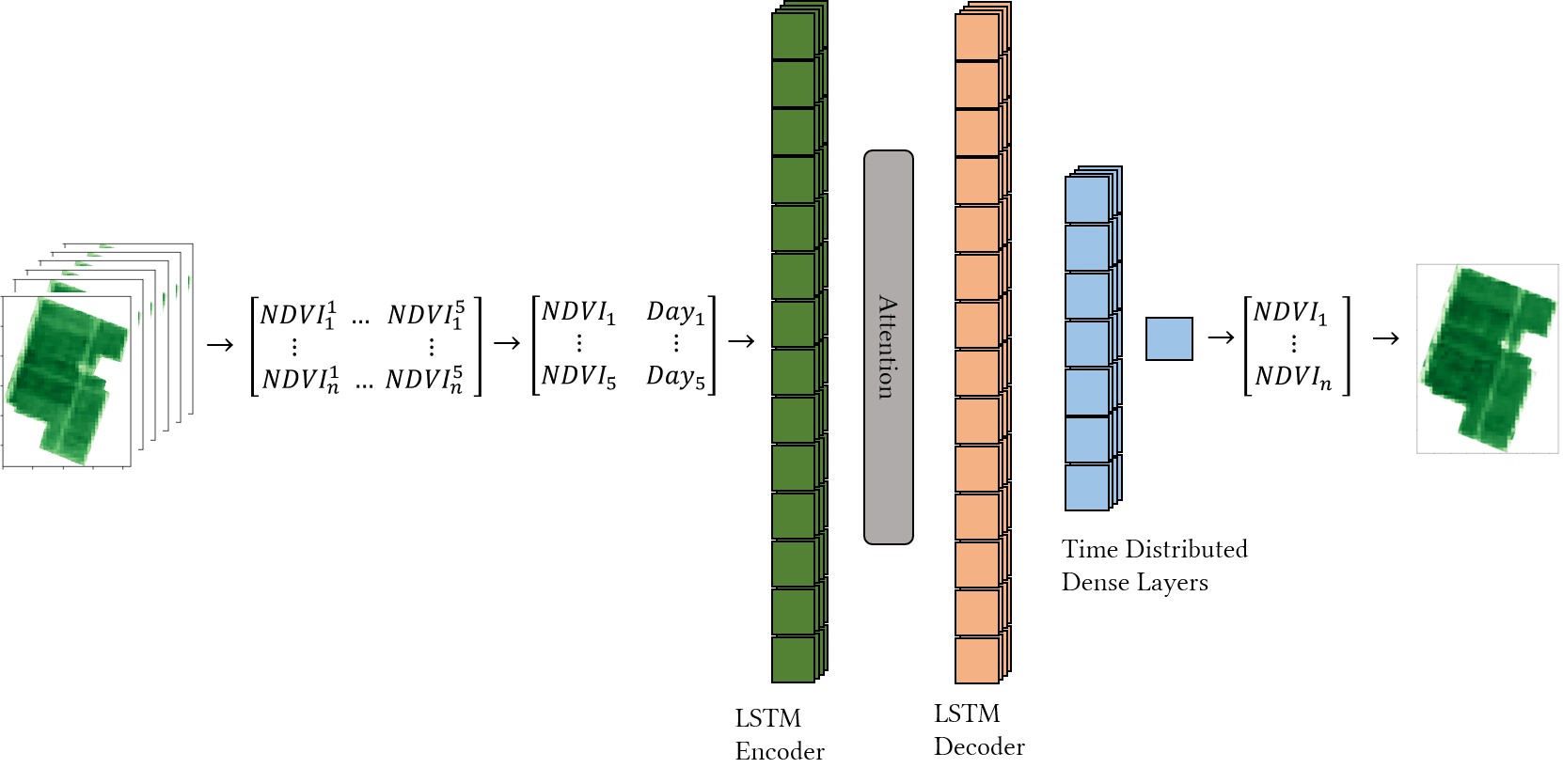}
    \caption{Structure of the LSTM encoder decoder model for prediction of NDVI time series}
    \label{fig:NDVI_lstm}
\end{figure}

These indices are per pixel calculations that are independent of the surrounding pixels meaning they can be translated into a simple time series prediction problem. The NDVI or NDWI images are converted into flattened matrices and then stacked to create a time series of each individual pixel through time. The model takes an input of the 5 previous NDVI images. Due to the irregular time interval of satellite imagery the 5 images were each coded with number of days from the target predicted image so that the output predictions are provided with a desired number of days in the future. The model again uses the sequence to sequence LSTM encoder-decoder structure \citep{Sutskever2014} with 50 neurons in each of the LSTM layers and then dense layers of 20 and 1 neurons. The output is a vector of the same length as the input with a single dimension representing one prediction per pixel. Once per pixel predictions are made on the latest imagery those predictions are reshaped into the structure of the initial image (Fig. \ref{fig:NDVI_lstm}). 

\subsection{Interpolation}
We used standard Gaussian kriging interpolation methods to determine soil moisture at all points between the sensors \citep{Yao2013} based on a linear distance function out from each sensor. We used soil moisture sensors and satellite imagery to supplement other features and provide a map to predict across. This was done on current data and then the outputs of the above predictions were used to produce future interpolation maps.

\section{Results and Discussion}

The LSTM predictions of soil moisture perform differently depending on the soil moisture depth that is being predicted. Generally, the average testing Root Mean Squared Error (RMSE) across the 14 days of prediction is highest at shallowest depths decreasing at the lower depths. It ranged between 2.4\% soil moisture error (value error) and 0.4\% soil moisture error with a mean of 1.23\% on soil moistures that vary between 15\% and 60\%. This outperforms previously published multiple linear regression models at 2\% \citep{Qiu2003} and is on par or outperforms \citet{Adeyemi2018}.

The LSTMs for NDVI and NDWI predictions were able to perform at a training RMSE of 0.027 and a testing RMSE of 0.065 on NDVIs that range between 0 and 1. The NDWI model has a training Mean Absolute Error (MAE) of  0.014 and a testing MAE of 0.02 on NDWI values that range between -0.3 to 0.55. 

We have observed variance in interpolation accuracy depending on the depth that was being predicted with the kriging scores (a proxy for accuracy varying between 0 and 1) between 0.82 – 0.97 with an average of 0.93 across all depths. 

\begin{figure}[h]
    \centering
    \includegraphics[width = 12cm]{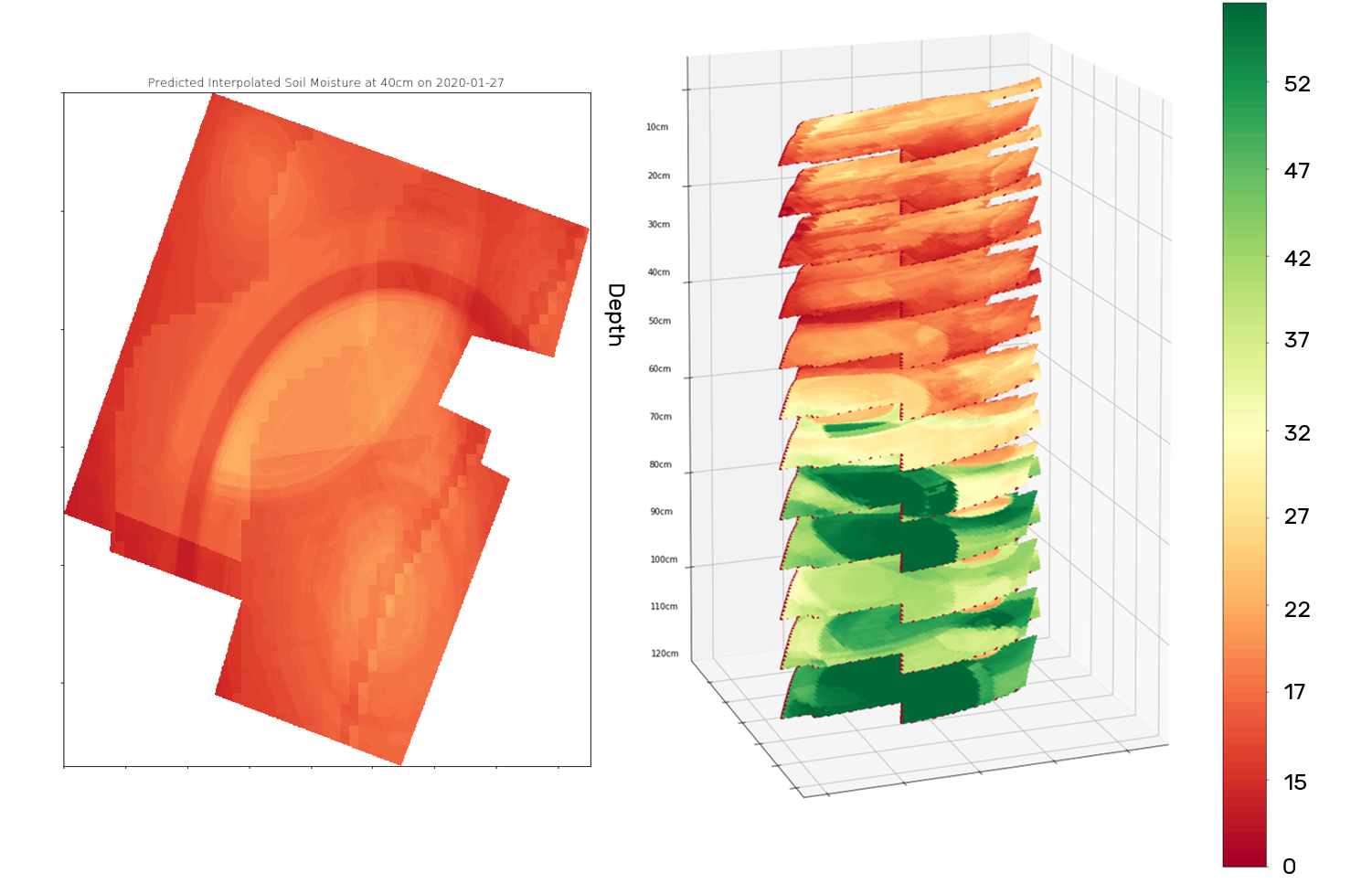}
    \caption{Left) Soil moisture interpolation across the satellite image at a single depth. Right) the interpolated soil moisture at all depths with each depth stacked on top of each other.}
    \label{fig:3d}
\end{figure}

The per pixel prediction maps are plotted in 3D across the X, Y and Z axes and coloured by soil moisture. Each soil moisture depth has its own predictive map produced and these are stacked so that depths can be viewed together (Fig.\ref{fig:3d}).

%\section{Discussion}
The potential to predict soil moisture across wide areas into the future can help agriculturalists on all scales. Informed knowledge of soil moisture can allow for more precise irrigation regimes or targeted irrigation only in areas that need it leading to savings in water usage. While water is an essential part of agriculture it is also required for drinking water and nearly all manufacturing processes, with climate change the available freshwater supply is expected to reduce leading to increased competition for fresh water \citep{Elliott2014} and new methods will be required for reducing water usage of all industries. Advanced warning of soil moistures that prevent a risk to crops can help increase yield, as the human population increases and with climate change some land becomes unusable for agriculture \citep{Zhang2011} or even human habitation increased crop yields will be required to maintain an adequate food supply. Precision agricultural regimes backed up by technology and artificial intelligence can work to make agricultural industries more sustainable.

\newpage

\bibliography{iclr2020_conference}

\begin{thebibliography}{23}
\providecommand{\natexlab}[1]{#1}
\providecommand{\url}[1]{\texttt{#1}}
\expandafter\ifx\csname urlstyle\endcsname\relax
  \providecommand{\doi}[1]{doi: #1}\else
  \providecommand{\doi}{doi: \begingroup \urlstyle{rm}\Url}\fi

\bibitem[Abbes et~al.(2019)Abbes, Magagi, and Goita]{Abbes2019}
Ali~Ben Abbes, Ramata Magagi, and Kalifa Goita.
\newblock {Soil Moisture Estimation from Smap Observations Using Long Short-
  Term Memory (LSTM)}.
\newblock In \emph{International Geoscience and Remote Sensing Symposium
  (IGARSS)}, pp.\  1590--1593. Institute of Electrical and Electronics
  Engineers Inc., jul 2019.
\newblock ISBN 9781538691540.
\newblock \doi{10.1109/IGARSS.2019.8898418}.

\bibitem[Adeyemi et~al.(2018)Adeyemi, Grove, Peets, Domun, and
  Norton]{Adeyemi2018}
Olutobi Adeyemi, Ivan Grove, Sven Peets, Yuvraj Domun, and Tomas Norton.
\newblock {Dynamic Neural Network Modelling of Soil Moisture Content for
  Predictive Irrigation Scheduling}.
\newblock \emph{Sensors}, 18\penalty0 (10):\penalty0 3408, oct 2018.
\newblock ISSN 1424-8220.
\newblock \doi{10.3390/s18103408}.
\newblock URL \url{http://www.mdpi.com/1424-8220/18/10/3408}.

\bibitem[Agrawal et~al.(2019)Agrawal, Barrington, Bromberg, Burge, Gazen, and
  Hickey]{Agrawal2019}
Shreya Agrawal, Luke Barrington, Carla Bromberg, John Burge, Cenk Gazen, and
  Jason Hickey.
\newblock {Machine Learning for Precipitation Nowcasting from Radar Images}.
\newblock dec 2019.
\newblock URL \url{http://arxiv.org/abs/1912.12132}.

\bibitem[Das \& Ghosh(2016)Das and Ghosh]{Das2016}
Monidipa Das and Soumya~K. Ghosh.
\newblock {Deep-STEP: A Deep Learning Approach for Spatiotemporal Prediction of
  Remote Sensing Data}.
\newblock \emph{IEEE Geoscience and Remote Sensing Letters}, 13\penalty0
  (12):\penalty0 1984--1988, dec 2016.
\newblock ISSN 1545598X.
\newblock \doi{10.1109/LGRS.2016.2619984}.

\bibitem[Efremova et~al.(2019)Efremova, Zausaev, and Antipov]{Efremova2019}
Natalia Efremova, Dmitry Zausaev, and Gleb Antipov.
\newblock {Prediction of Soil Moisture Content Based On Satellite Data and
  Sequence-to-Sequence Networks}.
\newblock jun 2019.
\newblock URL \url{http://arxiv.org/abs/1907.03697}.

\bibitem[Elliott et~al.(2014)Elliott, Deryng, M{\"{u}}ller, Frieler, Konzmann,
  Gerten, Glotter, Fl{\"{o}}rke, Wada, Best, Eisner, Fekete, Folberth, Foster,
  Gosling, Haddeland, Khabarov, Ludwig, Masaki, Olin, Rosenzweig, Ruane, Satoh,
  Schmid, Stacke, Tang, and Wisser]{Elliott2014}
Joshua Elliott, Delphine Deryng, Christoph M{\"{u}}ller, Katja Frieler, Markus
  Konzmann, Dieter Gerten, Michael Glotter, Martina Fl{\"{o}}rke, Yoshihide
  Wada, Neil Best, Stephanie Eisner, Bal{\'{a}}zs~M. Fekete, Christian
  Folberth, Ian Foster, Simon~N. Gosling, Ingjerd Haddeland, Nikolay Khabarov,
  Fulco Ludwig, Yoshimitsu Masaki, Stefan Olin, Cynthia Rosenzweig, Alex~C.
  Ruane, Yusuke Satoh, Erwin Schmid, Tobias Stacke, Qiuhong Tang, and Dominik
  Wisser.
\newblock {Constraints and potentials of future irrigation water availability
  on agricultural production under climate change}.
\newblock \emph{Proceedings of the National Academy of Sciences of the United
  States of America}, 111\penalty0 (9):\penalty0 3239--3244, mar 2014.
\newblock ISSN 00278424.
\newblock \doi{10.1073/pnas.1222474110}.

\bibitem[Frenken \& Gillet(2012)Frenken and Gillet]{Frenken2012}
K~Frenken and V~Gillet.
\newblock {Irrigation water requirement and water withdrawal by country}.
\newblock Technical report, Food and Agriculture Organisation of the United
  Nations, Rome, 2012.

\bibitem[Gao(1996)]{Gao1996}
Bo~Cai Gao.
\newblock {NDWI - A normalized difference water index for remote sensing of
  vegetation liquid water from space}.
\newblock \emph{Remote Sensing of Environment}, 58\penalty0 (3):\penalty0
  257--266, dec 1996.
\newblock ISSN 00344257.
\newblock \doi{10.1016/S0034-4257(96)00067-3}.

\bibitem[Hochreiter \& Schmidhuber(1997)Hochreiter and
  Schmidhuber]{Hochreiter1997}
Sepp Hochreiter and J{\"{u}}rgen Schmidhuber.
\newblock {Long Short-Term Memory}.
\newblock \emph{Neural Computation}, 9\penalty0 (8):\penalty0 1735--1780, nov
  1997.
\newblock ISSN 08997667.
\newblock \doi{10.1162/neco.1997.9.8.1735}.

\bibitem[Kingston et~al.(2009)Kingston, Todd, Taylor, Thompson, and
  Arnell]{Kingston2009}
Daniel~G. Kingston, Martin~C. Todd, Richard~G. Taylor, Julian~R. Thompson, and
  Nigel~W. Arnell.
\newblock {Uncertainty in the estimation of potential evapotranspiration under
  climate change}.
\newblock \emph{Geophysical Research Letters}, 36\penalty0 (20):\penalty0
  L20403, oct 2009.
\newblock ISSN 0094-8276.
\newblock \doi{10.1029/2009GL040267}.
\newblock URL \url{http://doi.wiley.com/10.1029/2009GL040267}.

\bibitem[{Le Hou{\'{e}}rou}(1996)]{LeHouerou1996}
Henry~N. {Le Hou{\'{e}}rou}.
\newblock {Climate change, drought and desertification}, 1996.
\newblock ISSN 01401963.

\bibitem[Li \& Heap(2011)Li and Heap]{Li2011}
Jin Li and Andrew~D. Heap.
\newblock {A review of comparative studies of spatial interpolation methods in
  environmental sciences: Performance and impact factors}, jul 2011.
\newblock ISSN 15749541.

\bibitem[Nay et~al.(2018)Nay, Burchfield, and Gilligan]{Nay2018}
John Nay, Emily Burchfield, and Jonathan Gilligan.
\newblock {A machine-learning approach to forecasting remotely sensed
  vegetation health}.
\newblock \emph{International Journal of Remote Sensing}, 39\penalty0
  (6):\penalty0 1800--1816, mar 2018.
\newblock ISSN 0143-1161.
\newblock \doi{10.1080/01431161.2017.1410296}.
\newblock URL
  \url{https://www.tandfonline.com/doi/full/10.1080/01431161.2017.1410296}.

\bibitem[Pagay et~al.(2016)Pagay, Kidman, and Jenkins]{Pagay2016}
Vinay Pagay, Catherine Kidman, and Allen Jenkins.
\newblock {Nitrogen and water status: Proximal and remote sensing tools for
  regionalscale characterisation of grapevine water and nitrogen status in
  Coonawarra}.
\newblock \emph{Wine {\&} Viticulture Journal}, 31\penalty0 (6):\penalty0
  42--47, 2016.
\newblock ISSN 1838-6547.

\bibitem[Qiu et~al.(2003)Qiu, Fu, Wang, and Chen]{Qiu2003}
Yang Qiu, Bojie Fu, Jun Wang, and Liding Chen.
\newblock {Spatiotemporal prediction of soil moisture content using
  multiple-linear regression in a small catchment of the Loess Plateau, China}.
\newblock \emph{Catena}, 54\penalty0 (1-2):\penalty0 173--195, nov 2003.
\newblock ISSN 03418162.
\newblock \doi{10.1016/S0341-8162(03)00064-X}.

\bibitem[Rolnick et~al.(2019)Rolnick, Donti, Kaack, Kochanski, Lacoste,
  Sankaran, Ross, Milojevic-Dupont, Jaques, Waldman-Brown, Luccioni, Maharaj,
  Sherwin, Mukkavilli, Kording, Gomes, Ng, Hassabis, Platt, Creutzig, Chayes,
  and Bengio]{Rolnick2019}
David Rolnick, Priya~L. Donti, Lynn~H. Kaack, Kelly Kochanski, Alexandre
  Lacoste, Kris Sankaran, Andrew~Slavin Ross, Nikola Milojevic-Dupont, Natasha
  Jaques, Anna Waldman-Brown, Alexandra Luccioni, Tegan Maharaj, Evan~D.
  Sherwin, S.~Karthik Mukkavilli, Konrad~P. Kording, Carla Gomes, Andrew~Y. Ng,
  Demis Hassabis, John~C. Platt, Felix Creutzig, Jennifer Chayes, and Yoshua
  Bengio.
\newblock {Tackling Climate Change with Machine Learning}.
\newblock jun 2019.
\newblock URL \url{http://arxiv.org/abs/1906.05433}.

\bibitem[Rouse et~al.(1973)Rouse, Hass, Schell, and Deering]{Rouse1973}
J.~W. Rouse, R.~H. Hass, J.A. Schell, and D.W. Deering.
\newblock {Monitoring vegetation systems in the great plains with ERTS}.
\newblock \emph{Third Earth Resources Technology Satellite (ERTS) symposium},
  1:\penalty0 309--317, jan 1973.
\newblock ISSN 00344257.
\newblock \doi{citeulike-article-id:12009708}.
\newblock URL \url{https://ntrs.nasa.gov/search.jsp?R=19740022614
  https://ntrs.nasa.gov/archive/nasa/casi.ntrs.nasa.gov/19740022614.pdf}.

\bibitem[Rumelhart et~al.(1986)Rumelhart, Hinton, and Williams]{Rumelhart1986}
David~E. Rumelhart, Geoffrey~E. Hinton, and Ronald~J. Williams.
\newblock {Learning representations by back-propagating errors}.
\newblock \emph{Nature}, 323\penalty0 (6088):\penalty0 533--536, 1986.
\newblock ISSN 00280836.
\newblock \doi{10.1038/323533a0}.

\bibitem[Stas et~al.(2016)Stas, {Van Orshoven}, Dong, Heremans, and
  Zhang]{Stas2016}
Michiel Stas, Jos {Van Orshoven}, Qinghan Dong, Stien Heremans, and Beier
  Zhang.
\newblock {A comparison of machine learning algorithms for regional wheat yield
  prediction using NDVI time series of SPOT-VGT}.
\newblock In \emph{2016 5th International Conference on Agro-Geoinformatics,
  Agro-Geoinformatics 2016}. Institute of Electrical and Electronics Engineers
  Inc., sep 2016.
\newblock ISBN 9781509023509.
\newblock \doi{10.1109/Agro-Geoinformatics.2016.7577625}.

\bibitem[Sutskever et~al.(2014)Sutskever, Vinyals, and Le]{Sutskever2014}
Ilya Sutskever, Oriol Vinyals, and Quoc~V. Le.
\newblock {Sequence to Sequence Learning with Neural Networks}.
\newblock \emph{Advances in Neural Information Processing Systems}, 4\penalty0
  (January):\penalty0 3104--3112, sep 2014.
\newblock URL \url{http://arxiv.org/abs/1409.3215}.

\bibitem[Wahba(1990)]{Wahba1990}
Grace Wahba.
\newblock \emph{{Spline Models for Observational Data}}.
\newblock Society for Industrial and Applied Mathematics, jan 1990.
\newblock \doi{10.1137/1.9781611970128}.

\bibitem[Yao et~al.(2013)Yao, Fu, L{\"{u}}, Sun, Wang, and Liu]{Yao2013}
Xueling Yao, Bojie Fu, Yihe L{\"{u}}, Feixiang Sun, Shuai Wang, and Min Liu.
\newblock {Comparison of Four Spatial Interpolation Methods for Estimating Soil
  Moisture in a Complex Terrain Catchment}.
\newblock \emph{PLoS ONE}, 8\penalty0 (1), jan 2013.
\newblock ISSN 19326203.
\newblock \doi{10.1371/journal.pone.0054660}.

\bibitem[Zhang \& Cai(2011)Zhang and Cai]{Zhang2011}
Xiao Zhang and Ximing Cai.
\newblock {Climate change impacts on global agricultural land availability}.
\newblock \emph{Environmental Research Letters}, 6\penalty0 (1), 2011.
\newblock ISSN 17489326.
\newblock \doi{10.1088/1748-9326/6/1/014014}.

\end{thebibliography}
\bibliographystyle{iclr2020_conference}

\end{document}